\documentclass[useAMS,usenatbib]{mn2e}
\usepackage{rotating}
\usepackage{subfigure}
\RequirePackage{setspace}
\usepackage{graphicx}
\usepackage{fancyhdr}
\usepackage{natbib}
\usepackage{amsmath}
\usepackage{amssymb}
\usepackage{aas_macros}
\usepackage{epstopdf}

\title[Limits on the Molecular Gas Content of $z\sim5$ LBGs]{Limits on the Molecular Gas Content of $z\sim5$ LBGs}
\author[L. J. M. Davies et. al.]{L. J. M. Davies$^{1}$\thanks{E-mail:
Luke.Davies@bristol.ac.uk}, M. N. Bremer$^{1}$, E. R. Stanway$^{1}$,  M. Birkinshaw$^{1}$, M. D. Lehnert$^{2}$\\
$^{1}$Department of Physics, University of Bristol, H.H. Wills Physics Laboratory, Tyndall Avenue, Bristol, BS8 1TL, UK\\
$^{2}$Laboratoire d'Etudes des Galaxies, Etoiles, Physique et Instrumentation GEPI, Observatoire de Paris, UMR8111 du CNRS, Meudon, 92195 France}

\begin{document}

\date{Accepted: July 21 2010}

\pagerange{\pageref{firstpage}--\pageref{lastpage}} \pubyear{2010}

\maketitle

\begin{abstract}
  
  We present limits on the molecular gas content of Lyman Break
  Galaxies (LBGs) at $z\sim5$ from observations targetting redshifted
  CO(1-0) and CO(2-1) line emission. We observed a single field
  containing eight spectroscopically-confirmed $z\sim5$ LBGs, seven of
  which are contained within a narrow (z=4.95 $\pm$ 0.08) redshift
  range and the eighth is at $z=5.2$. No source was individually
  detected. Assuming the CO to H$_2$ conversion factor for
  vigorous starbursts, we place upper limits on the molecular gas
  content of individual $z\sim 5$ LBGs of M(H$_{2}$) $\lesssim
  10^{10}\,$M$_{\odot}$. From a stacking analysis combining all of the
  non-detections, the typical $z\sim5$ LBG has an H$_2$
  mass limit comparable to their stellar mass, $< 3.1 \times
  10^{9}$\,M$_{\odot}$. This limit implies that, given the star
  formation rates of these systems (measured from their UV emission),
  star formation could be sustained for at most $\sim 100$\,Myr, similar
  to the typical ages of their stellar populations. The lack of a
  substantially larger reservoir of cold gas argues against the LBGs
  being UV luminous super starbursts embedded in much larger UV-dark
  systems and as a result increases the likelihood that at least those
  LBGs with multiple components are starbursts triggered by
  mergers. The sources responsible for reionization are expected to be
  starbursts similar to these systems, but with lower luminosities,
  masses and consequently with star formation timescales far shorter
  than the recombination timescale. If so, the ionized bubbles
  expected in the IGM during the reionization era will only infrequently
  have UV-luminous sources at their centres.

\end{abstract}


\section{Introduction}
\label{Into}

Detailed observations of the earliest galaxies are necessary if we are
to form a complete picture of galaxy formation and evolution. While
increasing numbers of spectroscopically-confirmed galaxies are being
discovered at $z\sim 5$ and above (within $\sim$1Gyr of the Big
Bang), almost all are discovered through rest-frame UV emission
originating from strong ongoing star formation. Unfortunately, this
tells us little about the (potentially dominant) UV-dark baryonic
component of these galaxies and consequently limits our understanding of
star formation in the high-$z$ universe.

Lyman-break Galaxies (LBGs) form a substantial part of current samples
of $z>5$ galaxies
\citep[\textit{e.g.}][]{vanzella09,Douglas09,Douglas10}. They are
identified \textit{via} their UV continuum emission longward of 1216\AA~ in
the rest-frame, which arises from hot young stars formed in unobscured
starburst regions. While {\it Spitzer}-based follow-up observations
have made some progress in exploring and placing limits on their older
underlying stellar populations
\citep[\textit{e.g.}][]{eyles07,Verma07}, we know little about their
gas content, which is a crucial diagnostic of the duration of the
ongoing starburst and of the nature of the LBGs themselves. The
galaxies have unobscured star formation rates of a few $\times
\,10$\,M$_\odot$\,yr$^{-1}$ arising from regions with a typical surface area
of $\sim 1$\,kpc$^2$. The strong wind generated by such a starburst can
potentially limit the available fuel for continuing star
formation. Clearly, the amount of gas available to the ongoing
starburst is key to an understanding of the nature of the burst and the future
of star formation activity within the system. HST
imaging of $z\sim 5$ LBGs
\citep[\textit{e.g.}][]{Douglas10,conselice09} shows that many sytems
have multiple UV components and extended, distorted morphologies on
scales of a kpc or more. With the available optical and near-IR data, it
is currently impossible to determine whether such structures imply
that the LBGs originate in mergers, or are individual super starburst
regions embedded in much larger UV-dark systems, as found in low
redshift Lyman break analogues (\citet[]{overzier08}, see
\citet{Douglas10} for a discussion). If the latter scenario is correct,
then one would expect considerably more cool and cold gas 
present in the immediate environment of the LBGs than is present in
the stellar mass produced by the ongoing starbursts.

In this paper we probe the cool gas component of a sample of $z\sim 5$
LBGs drawn from the ESO Remote Galaxy Survey
\citep[ERGS\footnote{ESO Program ID: 175.A-0706},][]{Douglas07,Douglas09,Douglas10}. Two of the ten 40
arcmin$^{2}$ ERGS pointings display a large over-density of
spectroscopically-confirmed UV bright sources over narrow ($\Delta z
\sim 0.1$) redshift ranges. The first of these was the subject of a
previous letter \citep{Stanway08} in which we discussed the
identification of a UV-dark molecular line emitter at $z\sim5$.  Here,
we target one field, J1054.4-1245, which contains many
spectroscopically-confirmed LBGs in a two-arcminute diameter
region. As in our previous work, we target the field using the
Australia Telescope Compact Array (ATCA), but this time use the new
Compact Array Broadband Backend (CABB) which probes a $\Delta z\sim
0.6$ redshift range in a single exposure (at 38GHz). This is
approximately 15 times broader than in our first work and easily
encompasses the entire velocity range probed by the LBG over-density.

Throughout this paper, unless otherwise stated, all magnitudes are on the AB scale, and the cosmology used is \textit{H}$_{0}$=70kms$^{-1}$ Mpc$^{-1}$, $\Omega_{\Lambda}$ = 0.7 and $\Omega_{M}$ =0.3.

\section{Observations}                                             
\label{sec:obs}

\begin{figure}
\begin{center}

\includegraphics[scale=0.35, bb= 1 1 566 566]{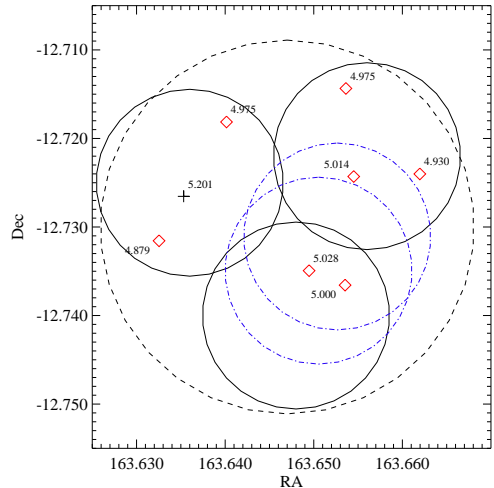}

\caption{The spatial position of spectroscopically-confirmed LBGs in our target field. Red diamonds indicate LBGs in the narrowest redshift over-density, while the black cross shows the eighth LBG in this region at slightly higher redshift. Our ATCA  pointings are indicated by the 7mm (solid circles for 2009 observations and dot-dashed blue circles for 2010 observations) and 12mm (dashed circle) half power beam widths. These encompass all eight LBGs.}

\label{fig:field_plot}
\end{center}
\end{figure}

Observations were carried out with the ATCA over two separate runs, one in
May 2009 as part of project C1954 and another in March 2010 as part of
project C2297. The primary goal of the project was to search for
UV-dark molecular line emitters across our target field and the
results of that study will be presented in a future paper. The array
was in the compact H168 configuration, using both North-South and
East-West baselines. We utilized the new CABB correlator with two
intermediate frequency (IF) bands, each with 2GHz/2048 channel
configurations.  We targeted CO(2-1) and CO(1-0) transitions at the
LBG over-density redshift. For CO(2-1) transitions we tuned the IF
bands to 36.70 GHz and 38.72 GHz to allow a simultaneous survey range
of $4.81\lesssim z \lesssim5.44$. We bin six adjacent 1 MHz channels
to increase signal to noise, and we obtain a spectral resolution of
$\sim$47\,km s$^{-1}$. During the first run, observations were taken
in six 8 hour periods between 2009 May 1 and 7.  Three pointings were
observed in order to target the maximum possible number of LBGs in the
field. During the second run two additional pointings were observed in
four 8 hour periods between 2010 March 21 and 23 (see figure
\ref{fig:field_plot}). A nearby bright source, PKS1054-188, was
observed every 15 minutes to determine the phase stability and the
pointing accuracy was was checked every hour. Primary flux calibration
was carried out on the standard ATCA calibration source, PKS1934-638,
each night. The half-power-beam-width (HPBW) of the ATCA at $\sim$37
GHz is 74$^{\prime\prime}$ and the restoring beam, for natural
weighting and this configuration, is 7.3$^{\prime\prime} \times
4.8^{\prime\prime}$.  For CO(1-0) transitions we tuned the IF bands
to 19.12 GHz and 21.12 GHz giving redshift coverage of $4.23\lesssim z
\lesssim5.34$. We observed a single pointing for two nights on 2009
May 8 to 9, encompassing all five $\sim$37GHz pointings (see Figure
\ref{fig:field_plot}). As before, we bin three adjacent 1 MHz channels
to increase signal to noise which gives a resolution of $\sim$47\,km
s$^{-1}$ (matching the velocity resolution of our high frequency
data), use nearby source 1054-188 for secondary flux calibration and
PKS1934-638 for primary flux calibration. The HPBW at $\sim$20 GHz is
2.5$^{\prime}$ and the restoring beam is 15.9$^{\prime\prime} \times
10.9^{\prime\prime}$. At $z\sim$5 this corresponds to a beam size of
$>50$kpc.

We expect any detectable line emission from our sources to have
velocity width $\sim$150\,km s$^{-1}$ \citep[the only two known high redshift,
non AGN, CO line-emitting galaxies at $z\sim5$ have line widths of
$\sim$160km s$^{-1}$, \citet{Coppin10} and $\sim$110km s$^{-1}$,
][]{Stanway08} and we require detections to be statistically
significant ($>2\sigma$) in each of 3 adjacent channels. Therefore by binning
our data to 3MHz and and 6MHz at the lower and higher frequency 
settings, a detection in three adjacent bins will correspond to a $\sim$150\,km s$^{-1}$-wide line.

Total integration times were 16 hours at each frequency giving an
rms noise of $\sim$0.11\,mJy/beam at 19 GHz and $\sim$0.17\,mJy/beam at
37GHz in each 47km s$^{-1}$ channel.

\begin{figure*}
\begin{center}
\includegraphics[scale=0.33]{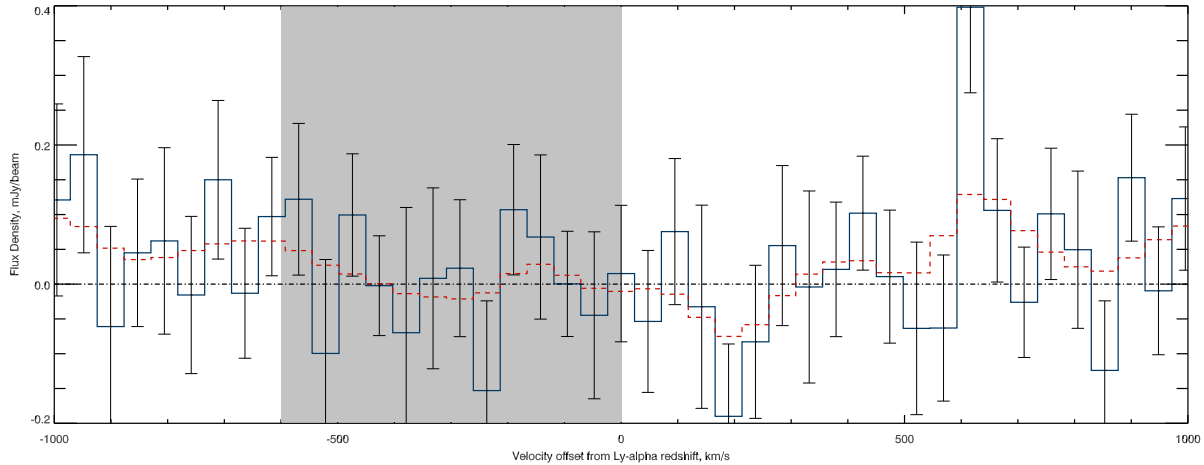}

\caption{1D spectrum of a typical LBG in our target region centered on the redshifted CO(2-1) line with redshifts assuming  the Ly-$\alpha$ redshift. The blue solid line is the target spectrum (binned to 6MHz channels) with errors displaying the 1$\sigma$ rms error in a $20^{\prime\prime}\,\times\,20^{\prime\prime}$ region around the LBG position. The red dashed line shows the target spectrum convolved with a 150km~s$^{-1}$ (FWHM) Gaussian. The shading highlights the region in which CO lines will found given typical offsets between Ly-$\alpha$ and systematic redshifts. The most significant positive peak in the above spectrum at +650km~s$^{-1}$  is still no more than 2$\sigma$ away from zero and consequently, given the number of independent bins displayed, is entirely consistent with noise.}

\label{fig:indu_LBG}
\end{center}
\end{figure*}    

\section{Limits on  LBG CO emission}
\label{indu_LBGs}

Our observations target eight spectroscopically-confirmed $z\sim 5$
LBGs which are spatially distributed within the half power beam widths
of individual pointings at both $\sim$37GHz and $\sim$19GHz. Seven lie
in the 3-dimensional over-density ($z\sim4.95\pm0.08$) of LBGs in the
larger ERGS field and the eighth is at slightly higher redshift
($z\sim5.2$). None of these galaxies show detectable line or continuum
emission at either of the redshifted CO transitions targeted.  For the
line emission, each extracted 1-dimensional spectrum was examined for
emission at the frequency expected for redshifted CO emission (using the
redshift determined from either Ly$\alpha$ emission or the break at
Ly$\alpha$ detected in optical spectroscopy
\citealt{Douglas10}). However, in a study of $z\sim 3$ star forming
galaxies, \citet{steidel10} show that the Ly$\alpha$ redshift can
differ from the systematic redshift derived from interstellar
absorption lines and other emission lines by up to $\sim
600$kms$^{-1}$, with the Ly$\alpha$ redshift being systematically
higher. Consequently, we additionally determined
limits to any line flux at frequencies corresponding to velocity
offsets of up to $600$kms$^{-1}$ shortward of the Ly$\alpha$-derived
value (see figure \ref{fig:indu_LBG}).

To search for CO line emission, 47 kms$^{-1}$-wide slices of the data
cube were extracted and the root-mean-square variation was ascertained
in each for a region of $\sim $20\,$\times$\,20 arcsec box centred on the
position of each LBG. Any real emission line is likely to be broader
than 50\,km~s$^{-1}$ as noted above. Consequently, we define a
detection as three consecutive 47 kms$^{-1}$ spectral bins at least
2$\sigma$ above zero flux in the region between 0 and 600 kms$^{-1}$
shortward of the Ly$\alpha$ redshift. Using these criteria, no
individual source was detected. To determine a characteristic line-limit
for each source, eleven values of the rms over three consecutive bins
in the same 0 to 600 kms$^{-1}$ region were obtained. As the
variation between these was small, we took the average value of these
as characteristic of a given source.  This results in a typical flux
limit of $<60$ mJy km~s$^{-1}$ for the CO(2-1) lines and $\lesssim40$
mJy km~s$^{-1}$ for the CO(1-0) line.  These correspond to luminosity
limits in each line of typically $1.1\times 10^{10}$ K km s$^{-1}$
pc$^{2}$ at $z\sim4.95$ for CO(2-1) and $2.6\times 10^{10}$ K km
s$^{-1}$ pc$^{2}$ for CO(1-0) \citep{Solomon97}. Converting this to a limit on the
molecular gas mass requires an appropriate conversion factor. In the
absence of a directly measured value at the highest redshifts, we use
the commonly-used conversion factor derived from local strongly star
forming galaxies \citep{Solomon05} of
M$_\mathrm{gas}$/L$^{\prime}_\mathrm{CO}=0.8$M$_{\odot}$ (K\,km~s$^{-1}$\,pc$^2$)$^{-1}$ given the
strength of the starbursts in these LBGs. Assuming that the emitting
medium is optically thick and thermalized, the line luminosity is independent of 
transition (\textit{i.e.}  L$^{\prime}_\mathrm{CO}$(2-1)=L$^{\prime}_\mathrm{CO}$(1-0), \cite{Solomon05}). The typical CO(2-1) limit places a constraint on the gas mass of M$_\mathrm{H_{2}} \lesssim 8.9
\times 10^{9}$ M$_{\odot}$. This is comparable to the stellar mass content
of LBGs at this redshift \citep[\textit{e.g.}][]{Verma07, Stark09}. The constraint placed by the
CO(1-0) line is a factor of two to four times higher depending upon
the source.

Of course, if the line width is much narrower than we assumed, we may
have missed emission using these criteria. However, we note that the
above flux and luminosity limits are appropriate for a 6$\sigma$ 
non-detection of any emission confined to a single channel. In fact,
no channel in the 600 kms$^{-1}$ range in any of the spectra deviates
from zero by more than 2.5$\sigma$, so the limit quoted above is
robust.

\begin{table*}
\centering

\begin{scriptsize}

\begin{tabular}{c c c c c c c c c}
\hline
\hline
RA&Dec&Redshift&I&R-I&I-$z$&rms$_\mathrm{CO(2-1)}^{1}$&L$^{\prime}_\mathrm{CO(2-1)}$$^{2}$&M$_{H_{2}}$ (CO(2-1))$^{3}$\\
(Deg)&(Deg)& &(mag)&(mag)&(mag) & (mJy/beam) & (x\,10$^{10}$\,K\,km~s$^{-1}$) & (x\,10$^{10}$\,M$_{\odot}$)\\
\hline

163.654&-12.7144&4.975$\pm$0.001&26.0$\pm$0.3&1.2$\pm$0.4&$<$-0.1&0.580&$<$1.28&$<$1.03 \\
163.640&-12.7181&4.975$\pm$0.001&25.8$\pm$0.2&1.4$\pm$0.4&$<$-0.3&0.595&$<$1.31&$<$1.05  \\
163.662&-12.7240&4.930$\pm$0.001&25.7$\pm$0.2&1.5$\pm$0.4& 0.5$\pm$0.3&0.530&$<$1.17&$<$0.94  \\
163.655&-12.7243&5.014$\pm$0.001&25.7$\pm$0.2&$>$2.3& 0.6$\pm$0.3&0.409&$<$0.90&$<$0.72  \\
163.635&-12.7265&5.201$\pm$0.001&26.1$\pm$0.3&$>$1.8&$<$0.0&0.583&$<$1.28&$<$1.02  \\
163.633&-12.7316&4.879$\pm$0.001&25.8$\pm$0.2&1.3$\pm$0.4&$<$-0.3&0.584&$<$1.29&$<$1.03  \\
163.649&-12.7349&5.028$\pm$0.001&26.2$\pm$0.3&$>$1.8&$<$0.1&0.334&$<$0.74&$<$0.59 \\
163.654&-12.7366&5.000$\pm$0.001&26.1$\pm$0.3&1.3$\pm$0.5&$<$-0.0&0.348&$<$0.77&$<$0.61   \\

\\
 
\hline
\end{tabular}
\end{scriptsize}
\caption{Properties of LBGs in target field (Positions given in J2000). \textbf{NOTES -} $^{1}$Average of the total of the 2$\sigma$ rms errors in 3 adjacent bins (total in 150km\,s$^{-1}$ width) for all bins, 600km\,s$^{-1}$ blue-ward of the CO(2-1) line position if optical redshifts are correct.
$^{2}$Limit of CO luminosities derived from rms errors for an unresolved $\sim$150km~s$^{-1}$ line (three channels at 2$\sigma$ rms limit).
$^{3}$Inferred gas mass derived from conversion factor for local infrared-luminous galaxies \citep{Solomon05}. }

\label{tab:gas_mass}        
\end{table*}

In addition to determining limits on individual sources, we can derive
deeper limits on the ``typical'' $z\sim 5$ LBG by creating an average
stack from the spectra of the eight individual sources, having shifted
each spectrum to the same effective redshift.  Concentrating on the
CO(2-1) data and using the same criteria as for an individual
spectrum, no line was detected in the averaged spectrum. The limits on
the flux, line luminosity and H$_2$ mass of a ``typical'' LBG are tightened
to  $S_{\mathrm{CO}} \Delta \nu \lesssim17.5$ mJy km~s$^{-1}$, $L^{'}_{\mathrm{CO(2-1)}} < 3.9\times 10^9$ K km
s$^{-1}$ pc$^{2}$ and M$_\mathrm{H_{2}} \lesssim 3.1 \times 10^{9}$ M$_{\odot}$ at
$z=4.95$ (2$\sigma$).

\begin{figure}
\begin{center}
\includegraphics[scale=0.35]{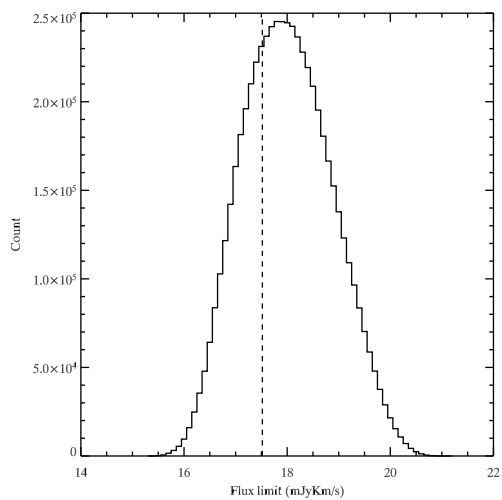}

\caption{Histogram of the distribution of flux limits from 823543 possible iterations of stacking LBG spectra. Eight LBG spectra were combined varying the relative offset in redshift between seven of the  spectra in six steps,  +150, +100, +50, 0, -50, -100 and -150 kms$^{-1}$ relative to the eighth spectrum. The dashed line shows the  flux limit for the eight spectra stacked at the optically-derived redshifts. The distribution of flux limits is Gaussian-distributed as expected given the noise characteristics of individual velocity channels. }

\label{fig_limits}
\end{center}
\end{figure}

However, this process is not completely straightforward as the true
redshifts for the sources may not exactly match the Ly$\alpha$-derived
values, as noted above.  The offsets between the Ly$\alpha$ and true
values may vary by $\sim 300$km~s$^{-1}$ between objects. Consequently, any
combination of weak lines may be averaged in such a way that they are
smeared out in frequency and do not reinforce each other to become a
detection in the average spectrum. To explore this we made multiple
average spectra, varying the relative offset in redshift between each
spectrum in seven steps, +150, +100, +50, 0, -50, -100 and -150
kms$^{-1}$, in total $7^7$ or 823543 average spectra. We then take the
ten flux limits in the -600 to 0km~s$^{-1}$ region from each spectrum.
The distribution of flux limits for these spectra are shown in figure
\ref{fig_limits}.  As can be seen, using the Ly$\alpha$ redshifts does
not significantly underestimate the true value of the limit.

For completeness, we carried out the same procedure, but looking at
features contained within a single bin as would be the case for
extremely narrow lines.  The highest flux in one of the -600 to 0
kms$^{-1}$ bins in the average spectra was 12.6 mJy kms$^{-1}$
(consistent with the noise characteristics given the 823543 different
realisations), considerably less than the above limit of
$S_{\mathrm{CO}} \Delta \nu \lesssim17.5$ mJy kms$^{-1}$ and so we
consider that limit to be robust even for a very narrow line.

\section{Discussion}

Assuming that the conversion between CO luminosity and H$_2$ mass
determined by \citet{Solomon05} is applicable (the intensity of the
starbursts in these LBG indicates that it is), the non-detection of CO
emission from these LBGs places interesting constraints on their
nature.  The similarity between the typical stellar masses of such
systems as determined from multiband photometry of $z\sim 5$ LBGs
\citep{Verma07}, and the limit to their typical H$_2$ mass (a few
times larger) would indicate that the systems are observed about half
way through the ongoing burst of star formation assuming that the
current star formation rate is maintained and the conversions from gas
to stars approaches $\sim 100$ per cent efficiency. However,
the efficiency with which gas is converted into stars is typically no more 
than ten per cent, and potentially only one or two per cent \citep{Lehnert09}, thus suggesting
that these sources are much more than halfway through their life
cycle. This is unsurprising as timescales of $\sim$ 10 Myrs will be
required to produce and sustain the observed UV continuum fluxes from
a population of O and B stars. Dividing the limiting gas mass by the
typical star formation rate from \citet{Verma07} gives a timescale of
$\sim 100 $Myr, with the typical age of a stellar population
determined by \citet{Verma07} being a few tens of Myr. As this
implicitly assumes an unrealistic complete transformation of gas into
stars, even if the CO luminosity to H$_2$ conversion factor is higher
than the Solomon \& Vanden Bout value, this is a robust upper limit to
the star formation timescale at this level of activity.

The apparent lack of a comparatively large reservoir of molecular gas
in and around the UV-luminous system strongly argues against the LBGs
being unobscured super starbursts embedded in much more massive and
extended underlying systems. Any such system should contain
sufficiently large amounts of enriched molecular gas to be detectable
here. Consequently, it provides support for the alternative hypothesis
of a merger origin for at least those systems with multiple
spatially-distributed UV luminous components. If all $z\sim 5$ LBGs
are triggered by mergers with other galaxies, our lack of a single
detection implies that few involve comparatively massive systems with
substantial obscured star formation or large amounts of processed
molecular gas. Additionally, the most active phases of a merger occur
when the gas has been consumed and is concentrated in the central
regions of the systems. If LBGs are triggered by mergers, and given
the need to build UV continuum fluxes through star formation, this may
take substantial time. Therefore in this model LBGs may only be
detectable (in the rest frame UV) towards the end of their starbursts. Observed luminous
starburst galaxies at high-$z$ typically have short gas
depletion times relative to their ages, similar to that seen in these
sources. The lack of detectable 850 micron continuum flux in APEX/LABOCA observations
of different but essentially similar LBGs \citep{Stanway10} gives further support to this model. 
As noted in \cite{Stanway10} the conversions from continuum flux limits to star formation rates are 
currently ambiguous. However, the lack of detectable
dust emission from these similar sources lends support to the idea that
 they do not have large reservoirs of cool and cold material, thus strengthening 
 the idea that these sources are not embedded in more massive obscured systems.

There has been  recent discussion
\citep{papadopoulos10,gnedin10} of whether there should be 
deviation away from the low-redshift Schmidt-Kennicutt (SK) relation
\citep{schmidt59,kennicutt98} at high redshift. Here we note that for
unobscured star formation rates of a few tens of solar masses per year
arising from regions of approximately $\sim 1$kpc \citep[\textit{e.g.}][]{Verma07, Bouwens04, Bremer04}, a
limit of M$_\mathrm{H_{2}} \lesssim 3.1 \times 10^{9}$ M$_{\odot}$ in the
same area is consistent with the low-redshift SK relation. The synthesised
ATCA beam at 38GHz probes an area of well over 100 kpc$^2$ centred on each
LBG and so if there was an underlying larger UV-dark galaxy within
which each LBG was embedded, any significant obscured star formation
in that galaxy would lead to deviation from the nearby SK relation. As
simulations imply that the relation gets steeper with redshift
\citep[more gas per unit star formation, \textit{e.g.}][]{gnedin10},
the deviation from the SK relation would only increase if the nearby
relation is inappropriate.

Given the comparatively brief elapsed time between the end of
reionization and the redshift explored here, the limited gas
reservoirs available for star formation in these systems and their
relatively short star formation timescales have an important
consequence for the reionization process. Reionization is likely to be
dominated by as yet undetected faint, low mass UV-luminous starbursts
\citep[\textit{e.g.}][]{Lehnert03,Salvaterra10}. As a consequence of
their lower mass, these sources will have even shorter star formation
lifetimes than the LBGs observed here. At all redshifts the
recombination time for the IGM is longer than the Hubble time and
hence also much longer than the UV-luminous lifetime of a low mass
starburst. Consequently, most of the ionized bubbles expected to exist
in the otherwise neutral IGM during reionization will only
occasionally have UV-luminous sources at their centres, and most
starbursts giving rise to the ionising flux will have long-since
ceased to actively form stars and therefore have faded in the UV.

\section*{Acknowledgements}

LJMD and ERS gratefully acknowledge support from the UK Science and Technology 
Facilities council. Based on data from ATCA programs C1954 and C2297. The Australia 
Telescope Compact Array is part of the Australia Telescope which is funded by the 
Commonwealth of Australia for operation as a National Facility managed by CSIRO. The 
authors thank M. Voronkov and R. Ward for their assistance as Duty Astronomers
at the ATCA.

\end{document}